\documentclass{physeauth}
\usepackage{graphicx}
\usepackage{amsmath}
\usepackage{amssymb}

\begin{document}

\begin{frontmatter}

\title{Entanglement evolution in a cascaded system with losses} 

\author{Christian Di Fidio},
\author{Werner Vogel \thanksref{thank1}}

\address{Arbeitsgruppe Quantenoptik, Institut f\"ur Physik,
Universit\"at Rostock, D-18051 Rostock, Germany}

\thanks[thank1]{Corresponding author.~E-mail:~werner.vogel@uni-rostock.de}

\begin{abstract}

The dynamics of a cascaded system that
consists of two atom-cavity subsystems 
is studied by using the quantum trajectory method.
Unwanted losses are included, such as photon absorption
and scattering by the cavity mirrors and spontaneous emission
of the atoms.
Considering an initially excited two-level atom in 
the source subsystem, analytical solutions 
are obtained.
The entanglement evolution is studied for the two atoms
and for the two intracavity fields.

\end{abstract}

\end{frontmatter}


\section{Introduction}
The concept of entanglement has been of great interest
since the early days of quantum mechanics~\cite{Schroedinger}, 
and it has become of central importance in a variety of 
discussions on the fundamental aspects of the 
theory~\cite{Einstein,Bell}. 
Unexpectedly, entanglement has also been
demonstrated in classical Brownian motion~\cite{AKN05}.
Nowadays
entanglement is receiving new attention in the
rapidly developing fields of quantum information,
quantum computation and quantum technology; for 
reviews, see~\cite{Nielsen,HarocheRaimond}.
In the context of entanglement preparation between
atoms at separate nodes, a variety of schemes have 
been proposed, for example,
by measuring the superpositions of light fields 
released from
separate atomic samples or 
by measuring a probe light field that 
has interacted in a prescribed way with different samples.
Due to the indistinguishability in the measurement
and conditioned on the results of the measurements,
the atomic system is projected onto an entangled 
state~\cite{Bose:5158,Duan:5643,Duan:253601}.
An unconditional preparation of entanglement
has also been analyzed in the case of a cascaded system.
This has been discussed for
two distantly separated atoms~\cite{Clark:177901,Gu:043813},
as well as for separate atomic ensembles~\cite{Parkins:053602}.
More recently, the entanglement evolution
for a Raman-driven cascaded system has also 
been analyzed~\cite{Difidio:032334}.

In the spirit of these previous achievements, 
in the present contribution
we will consider 
a cascaded open quantum system. 
We study the dynamics of a system that consists of two
atom-cavity sub-systems $A$ and $B$.
The quantum source $A$ emits a photon and the second
quantum subsystem $B$ reacts on the emitted photon.
Unwanted losses are included, such as photon absorption
and scattering by the cavity mirrors and spontaneous emission
of the atoms.
Considering an initially excited two-level atom in 
the source subsystem, analytical solutions 
are obtained.
Subsequently, the entanglement evolution between the two atoms,
as well as between the two intracavity fields,
is studied by using the concurrence.

The paper is organized as follows.
In Sec.~\ref{section2} the master equation 
describing the dynamics of the cascaded
system is introduced,
and the problem is solved
analytically by using the quantum trajectory method.
In Sec.~\ref{section3} 
the entanglement evolution
between the two atoms, or between the two intracavity fields, is analyzed. 
Finally, some concluding remarks are given in Sec.~\ref{conclusions}.

\section{Cascaded system dynamics}
\label{section2}

\begin{figure}[h]
\includegraphics[width=7.0cm]{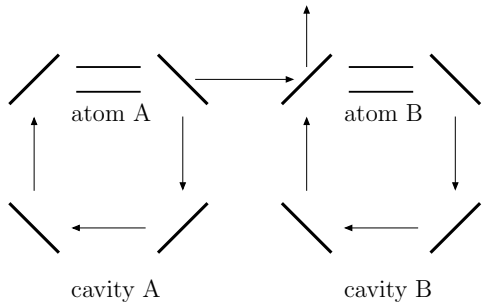}
\caption{The cascaded open system
consisting of two atom-cavity subsystems 
$A$ and $B$.}
\label{fig:figure_pra_1}
\end{figure}

In this section we analyze the dynamics of 
the system under study.
The cascaded open quantum system consists of two
atom-cavity subsystems $A$ and $B$,
where the source subsystem $A$ is cascaded with the
target subsystem $B$, cf. Fig.~\ref{fig:figure_pra_1}.
The cavities have three perfectly reflecting mirrors and one
mirror with transmission coefficient $T \ll 1$.
In the two subsystems $A$ and $B$
we consider a 
two-level atomic transition
of frequency  $\omega_k$ (related to the atomic
energy eigenstates $|1_k\rangle$ and $|0_k\rangle$)
coupled to a cavity mode of frequency 
$\omega_k'$, where $k=a,b$ denotes the subsystem.
The cavity mode is detuned by $\Delta_k$
from the two-level atomic transition frequency, 
$\omega_k = \omega_k' + \Delta_k$, 
and  is damped by losses through the partially transmitting
cavity mirrors. 
In addition to the wanted outcoupling of the field,
the photon can be spontaneously emitted out the side
of the cavity into modes other than the one
which is preferentially coupled to the resonator.
Moreover, the photon may be absorbed or scattered
by the cavity mirrors.
It has been shown that unwanted losses can have significant
effects on the dynamical evolution of a quantum system
and cannot be a priori neglected,
see, e.g., Refs.~\cite{Difidio:031802,Difidio}.

To describe the dynamics of the system we 
will use a master equation formalism.
This leads 
to the following master equation for the reduced density 
operator $\hat \rho(t)$ of the system:
\begin{eqnarray}
\frac{d\hat\rho(t)}{dt} &=& \frac{1}{i\hbar} \! \left[\hat H,\hat\rho(t)\right] \!+\! \sum_{i = 1}^5
\left[ \hat J_i \hat\rho(t) \hat J_i^\dagger 
- \frac{1}{2} \hat J_i^\dagger \hat J_i \hat\rho(t) \right. \nonumber \\
&-& \left. \frac{1}{2} \hat\rho(t) \hat J_i^\dagger \hat J_i \right] .
\label{eq:master_1}
\end{eqnarray}
The Hamiltonian is given by
\begin{equation}
\hat H = \hat H_A + \hat H_B + i \hbar \frac{\sqrt{\kappa_a \kappa_b}}{2}
\left( e^{-i \phi} \hat b \hat a^\dagger - e^{i \phi} \hat b^\dagger \hat
a \right) \, ,
\label{eq:hamiltonian}
\end{equation}
where $\hat H_A$ and $\hat H_B$ describe the atom-cavity interaction in the two subsystems $A$ and $B$, respectively,
and, in the
rotating-wave approximation, are given by
\begin{equation}
\hat H_A = \hbar g_a  \left( \hat{a} \hat A_{10} 
+ \hat a^\dagger \hat A_{01}\right) + \hbar \Delta_a {\hat A}_{11}
   \, ,
\label{eq:JC_hamiltonian_A}
\end{equation}
and
\begin{equation}
\hat H_B = \hbar g_b \left( \hat{b} \hat B_{10} 
+ \hat b^\dagger \hat B_{01}\right) + \hbar \Delta_b {\hat B}_{11}
   \, .
\label{eq:JC_hamiltonian_B}
\end{equation}
The third term in Eq.~(\ref{eq:hamiltonian}) describes the
coupling between the two
cavities~\cite{Carmichael:2273,Gardiner:2269}. 
In these expressions, $\hat a$ ($\hat a^\dagger$) 
is the annihilation (creation) operator
for the cavity field $A$
and similarly $\hat b$ ($\hat b^\dagger$) for the cavity field $B$.
We have also defined
$\hat A_{ij} = |i_a\rangle \langle j_a|$ ($i,j = 0, 1$),
and $\hat B_{ij} = |i_b\rangle \langle j_b|$ ($i,j = 0, 1$).
In addition, $g_k$ is the atom-cavity coupling
constant and $\kappa_k$ the cavity bandwidth. 
The phase $\phi$ is related to the phase change upon reflection from the source output mirror,
and/or to the retardation of the source due
to the spatial separation between the source and the target,
cf.~\cite{Carmichael2}.

The jump operators $\hat J_i$ entering in Eq.~(\ref{eq:master_1}) are defined by
\begin{equation}
\hat J_1 = \sqrt{\kappa_a} \hat a + \sqrt{\kappa_b} e^{-i\phi}
\hat b \, ,
\label{jump_1}
\end{equation}
which describes photon emission by the cavities;
\begin{equation}
\hat J_2 = \sqrt{\kappa_a'} \hat a \, , ~~~~~~~
\hat J_3 = \sqrt{\kappa_b'} \hat b \, ,
\label{jumps_2_3}
\end{equation}
are associated with photon absorption or scattering
by the cavity mirrors; and
\begin{equation}
\hat J_4 = \sqrt{\Gamma_a} \hat A_{01} \, , ~~~~~~~
\hat J_5 = \sqrt{\Gamma_b} \hat B_{01} \, ,
\label{jumps_4_5}
\end{equation}
are related to a photon spontaneously emitted by the atoms.
Here $\kappa_k'$ and $\Gamma_k$
are the cavity mirrors' absorption (or scattering) rate
and the spontaneous 
emission rate of the two-level atom, respectively.
Note that the operator $\hat J_1$ contains 
the superposition of the two fields radiated by the
two cavities
due to the fact that the radiated photon cannot be associated with 
photon emission from either $A$ or $B$ separately.

In the following we will identify, for notational convenience,
the state $|a\rangle$ with the state $|1,0,0,0\rangle$, 
which denotes the atom $A$ in
the state $|1_a\rangle$, the cavity $A$ in the vacuum
state, the atom $B$ in the state $|0_b\rangle$ , and
the cavity $B$ in the vacuum state.
In the state $|b\rangle \equiv |0,1,0,0\rangle$
the atom $A$ is in
the state $|0_a\rangle$, and the cavity $A$ is in the 
one-photon Fock state.
Similarly, we define $|c\rangle \equiv |0,0,1,0\rangle$,
$|d\rangle \equiv |0,0,0,1\rangle$, and
$|e\rangle \equiv |0,0,0,0\rangle$.
The state $|a\rangle$ will be considered as the initial
state of the system. It follows that the Hilbert space that describes the cascaded system under study is, in our 
model, spanned by the five state vectors $|a\rangle$, $|b\rangle$, $|c\rangle$, $|d\rangle$, and $|e\rangle$.

To evaluate the time evolution of the system 
we use a quantum trajectory
approach~\cite{Carmichael2,Dalibard:580,Dum:4382}.
Let us consider the system prepared at time
$t_0 = 0$ in the state $|a\rangle$. To determine 
the state vector of the system at a later time $t$, 
assuming that no jump has occurred between time $t_0$ and $t$,  
we have to solve the nonunitary Schr\"{o}dinger
equation
\begin{equation}
i \hbar \frac{d}{dt} 
| \bar{\psi}_{\rm no} (t) \rangle  =\hat{H^{'}} \, | \bar{\psi}_{\rm no} (t) \rangle \, ,
\label{eq:schr_non_unitary} 
\end{equation}
where $\hat{H^{'}}$ is the non-Hermitian Hamiltonian
given by
\begin{eqnarray}
\hat{H^{'}} &=& \hat H
- \frac{i \hbar}{2} \sum_{i=1}^5 \hat J_i^\dagger
\hat J_i = \hat H_A + \hat H_B - i \hbar \Big(
\frac{K_a}{2}
\hat a^\dagger \hat a  \nonumber \\
&+& \frac{K_b}{2} \hat b^\dagger \hat b
+ \frac{\Gamma_a}{2} \hat A_{11} + 
\frac{\Gamma_b}{2} \hat B_{11} + 
\sqrt{\kappa_a \kappa_b} e^{i \phi} \hat b^\dagger \hat a \Big) ,
\label{eq:n_H_Hamiltonian}
\end{eqnarray}
where we have defined
\begin{equation}
K_a = \kappa_a + \kappa_a' \, , ~~~~
K_b = \kappa_b + \kappa_b' \, .
\label{eq:kappa}
\end{equation}
If no jump has occurred between time $t_0$ and $t$, the system evolves via Eq.~(\ref{eq:schr_non_unitary}) into the unnormalized state
\begin{equation}
| \bar{\psi}_{\rm no} (t) \rangle = \alpha(t) |a\rangle + \beta(t) |b \rangle + \gamma(t) |c\rangle + \delta(t) |d \rangle \, .
\label{eq:psi_n}
\end{equation}
In this case the conditioned density operator for
the atom-cavity system is given by 
\begin{equation}
\hat \rho_{\rm no} (t) = \frac{ | \bar{\psi}_{\rm no} (t) \rangle \langle
\bar{\psi}_{\rm no} (t) |}{\langle \bar{\psi}_{\rm no} (t) |\bar{\psi}_{\rm no} (t) \rangle}
\, ,
\label{eq:rho_no}
\end{equation}
where we indicate with ``no" the fact that
no jump has occurred between time $t_0$ and $t$.

The evolution governed by the nonunitary Schr\"odinger equation~(\ref{eq:schr_non_unitary})
is randomly interrupted by one of the five 
kinds of jumps $\hat J_i$,
cf. Eqs.~(\ref{jump_1})-(\ref{jumps_4_5}).
If a jump has occurred at time
$t_{\rm J}$, $t_{\rm J} \in (t_0, t]$, the
wave vector is found collapsed into the state $|e \rangle$ due to the action of one of the jump operators
\begin{equation}
\hspace*{-0.5cm} \hat J_{i} \, | \bar{\psi}_{\rm no} (t_{\rm J}) \rangle  \rightarrow |e \rangle   ~~(i = 1, \ldots, 5). \label{eq:jump_op_i} 
\end{equation}
In the problem under study we may have only one jump. Once the system collapses into the state $|e\rangle$, the nonunitary Schr\"odinger equation~(\ref{eq:schr_non_unitary}) 
lets it remain unchanged.
In this case the conditioned density operator at time $t$ is given by
\begin{equation}
\hat \rho_{\rm yes} (t) = | e \rangle \langle e | \, ,
\label{eq:rho_yes}
\end{equation}
where we indicate with ``yes" the fact that a jump has occurred.

In the quantum trajectory method, the density operator $\hat \rho (t)$ is obtained by performing an ensemble
average over the different conditioned density operators at
time $t$,
yielding the statistical mixture
\begin{equation}
\hat \rho (t) =  p_{\rm no}(t) \hat \rho_{\rm no} (t) +
  p_{\rm yes}(t) \hat \rho_{\rm yes} (t) \, .
\label{eq:rho_t}
\end{equation}
Here $p_{\rm no}(t)$ and $p_{\rm yes}(t)$ are the probability that between the initial time $t_0$ and time $t$ no jump and
one jump has occurred, respectively, where 
$p_{\rm no}(t) + p_{\rm yes}(t) = 1$.

To evaluate $p_{\rm no}(t)$ we use the method of the delay function~\cite{Dum:4382}. This yields the probability $p_{\rm no}(t)$
as the square of the norm
of the unnormalized state vector:
\begin{eqnarray}
p_{\rm no}(t) 
&=& \parallel | \bar{\psi}_{\rm no} (t) \rangle \! \parallel^2
= \langle \bar{\psi}_{\rm no} (t) |\bar{\psi}_{\rm no} (t) \rangle \nonumber \\ 
&=& |\alpha(t)|^2  + |\beta(t)|^2 + |\gamma(t)|^2  
+ |\delta(t)|^2\, .
\label{eq:p_no}
\end{eqnarray}
From Eqs.~(\ref{eq:rho_t}) and (\ref{eq:p_no}) one obtains for the density operator $\hat \rho (t)$ the expression
\begin{equation}
\hat \rho (t) =  
| \bar{\psi}_{\rm no} (t) \rangle \langle
\bar{\psi}_{\rm no} (t) |
+ |\epsilon(t)|^2 |e \rangle \langle e| \, ,
\label{eq:rho_t_1}
\end{equation}
where we have defined $|\epsilon(t)|^2 \equiv 1 - p_{\rm no}(t)$.
The physical meaning of $|\alpha(t)|^2$, $|\beta(t)|^2$, $|\gamma(t)|^2$, $|\delta(t)|^2$, and $|\epsilon(t)|^2$ 
is clear. They represent the probability that at time $t$ the system can be found either in $|a\rangle$,
$|b\rangle$, $|c\rangle$, $|d\rangle$, or $|e\rangle$.

In order to determine $\alpha(t)$, $\beta(t)$,
$\gamma(t)$, and $\delta(t)$, we have to solve
the nonunitary Schr\"odinger equation, cf. Eqs.~(\ref{eq:schr_non_unitary})
and~(\ref{eq:n_H_Hamiltonian}). 
This leads us to consider the following 
inhomogeneous system of differential equations, similar
to the one in Ref.~\cite{Difidio:032334},
\begin{equation}
\left\{ 
\begin{array}{llll}
\dot \alpha(t) 
= -i \left( \Delta_a - i \Gamma_a/2 \right)\alpha (t) - i g_a \beta(t) \, , \\
\dot \beta(t) = - i g_a \alpha(t) - (K_a/2) \beta(t) \, , \\
\dot \gamma(t) = -i \left( \Delta_b - i \Gamma_b/2 \right)\gamma (t) - i g_b \delta(t) \, ,  \\
\dot \delta(t) = - i g_b \gamma(t) - (K_b/2) \delta(t) 
-\sqrt{\kappa_a\kappa_b} e^{i \phi} \beta (t) \, .
\end{array}
\right.
\label{eq:sys_diff_eq_1}
\end{equation}
The differential equations for $\alpha(t)$ and
$\beta(t)$ can be solved independently from
those for $\gamma(t)$ and $\delta(t)$.
For the initial conditions $\alpha(0) \!=\!1$ and 
$\beta(0) \!=\! 0$, and defining
\begin{equation}
\Omega_k \!\equiv\! \sqrt{\frac{K_k^2}{4} \!-\! 4g_k^2 \!-\! i K_k \!\left(\Delta_k \!-\!i \frac{\Gamma_k}{2}\right) \!-\! \left(\Delta_k \!-\! i \frac{\Gamma_k}{2}\right)^2 } \, , 
\end{equation}
we can write the solutions as
\begin{eqnarray}
{\alpha} (t) &=&  \left[\frac{K_a/2 - i (\Delta_a - i \Gamma_a/2)}{\Omega_a} \sinh \left(\frac{\Omega_a t}{2}\right)  \right.
\nonumber \\
&+& \left. \cosh \left( \frac{\Omega_a t}{2}\right) \right] \! e^{-[(K_a +\Gamma_a)/4 
+ i \Delta_a/2]t} , \nonumber \\
{\beta} (t) &=& - \frac{2ig_a}{\Omega_a} \sinh \left(\frac{\Omega_a t}{2}\right) e^{-[(K_a + \Gamma_a)/4 + i\Delta_a/2]t}\, .
\label{eq:diff_eq_sol_general_a_b}
\end{eqnarray}
Inserting now in the inhomogeneous pair
of differential equations for $\gamma(t)$ and $\delta(t)$
the solution obtained for $\beta(t)$,
we can determine the solutions for $\gamma(t)$ and 
$\delta(t)$.
For the initial conditions
$\gamma(0) \!=\!0$ and $\delta(0) \!=\! 0$,
we can write the solutions as
\begin{eqnarray}
\hspace*{-0.0cm} {\gamma} (t) &=& g_b \left\{
f_{+}(t) \! \left[g_{-}(t) \!+\! h_{+}(t)  \right] 
\!-\! f_{-}(t) \!
\left[g_{+}(t) \!+\! h_{-}(t)  \right] \right\} ,
\nonumber \\
\hspace*{-0.0cm} {\delta} (t) &=& 
i \! \left[\frac{K_b \!-\! \Gamma_b}{4} \!-\! 
i \frac{\Delta_b}{2} \!+\! \frac{\Omega_b}{2}\right] \! f_{-}(t) 
\! \left[g_{+}(t) \!+\! h_{-}(t)  \right] \nonumber \\
&-& i \! \left[\frac{K_b \!-\! \Gamma_b}{4} \!-\!
i \frac{\Delta_b}{2} \!-\! \frac{\Omega_b}{2}\right] \! f_{+}(t) 
\! \left[g_{-}(t) \!+\! h_{+}(t)  \right] \!,
\label{eq:diff_eq_sol_general_g_d}
\end{eqnarray}
where we have defined, for notational convenience,
\begin{equation}
f_\pm(t) = \frac{g_a \sqrt{\kappa_a\kappa_b} e^{i \phi} }{\Omega_a \Omega_b} e^{[-(K_b + \Gamma_b)/4 - i \Delta_b/2 \pm \Omega_b/2]t} \, ,
\end{equation}
\begin{equation}
g_\pm(t) = \frac{e^{[(\Omega_a\pm \Omega_b)/2 - \Upsilon - i \Lambda]t}-1}{(\Omega_a\pm \Omega_b)/2 - \Upsilon - i \Lambda} \, ,
\end{equation}
and
\begin{equation}
h_\pm(t) = \frac{e^{-[(\Omega_a\pm \Omega_b)/2 + \Upsilon + i \Lambda]t}-1}{(\Omega_a\pm \Omega_b)/2 + \Upsilon + i \Lambda} \, ,
\end{equation}
where $\Upsilon \!=\! (K_a \!-\! K_b \!+\! \Gamma_a \!-\! \Gamma_b)/4$ and $\Lambda \!=\! (\Delta_a \!-\! \Delta_b)/2$.
\begin{figure}
\includegraphics[width=7.5cm]{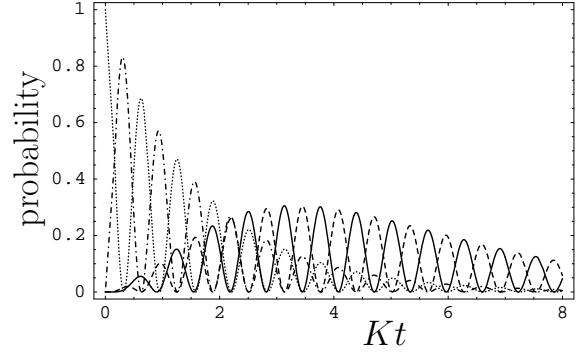}
\caption{The probabilities $|\alpha(t)|^2$ (dotted line), $|\beta(t)|^2$ (dot-dashed line), $|\gamma(t)|^2$ (full line), 
and $|\delta(t)|^2$ (dashed line) are shown for 
the case of equal parameters for the
two subsystems $A$ and $B$, where $g/K = 5$, $\kappa/K = 0.9$, $\Delta/K = 0.1$, $\Gamma/K= 0.2$.}
\label{fig:figure_pra_2}
\end{figure}
In the case of equal parameters for the
two subsystems $A$ and $B$, the solutions~(\ref{eq:diff_eq_sol_general_g_d}) simplify as 
\begin{eqnarray}
\gamma (t) &=& \frac{\kappa g^2 e^{i \phi}}{\Omega^3}
\left[e^{-\Omega t} \!+\! \Omega t \!-\! 1 \right] 
e^{[-(K + \Gamma)/4 - i \Delta/2 + \Omega/2]t}  \nonumber \\
&-& \frac{\kappa g^2 e^{i \phi}}{\Omega^3}
\left[e^{\Omega t} \!-\! \Omega t \!-\! 1 \right]
e^{[-(K + \Gamma)/4 - i \Delta/2 - \Omega/2]t} \, ,
\nonumber \\
\delta (t) &=& \frac{i\kappa g e^{i\phi}}{\Omega^3}
\left[\frac{K \!-\! \Gamma}{4} \!-\! i \frac{\Delta}{2}
\!+\! \frac{\Omega}{2}\right]\left[e^{\Omega t}
- \Omega t - 1 \right] \nonumber \\
&\times&
e^{[-(K + \Gamma)/4 - i \Delta/2 - \Omega/2]t} \!
- \frac{i\kappa g e^{i \phi}}{\Omega^3}\left[\frac{K \!-\! \Gamma}{4} \!-\! i \frac{\Delta}{2} \right.\nonumber \\
&-& \left. \frac{\Omega}{2}\right]\left[e^{-\Omega t} \!+\! \Omega t \!-\! 1 \right]  e^{[-(K + \Gamma)/4 - i \Delta/2 + \Omega/2]t} .
\label{eq:diff_eq_sol_equal_values}
\end{eqnarray}
where we have used
$\lim_{x \to 0} \{[\exp(\pm x t) - 1]/x\} = \pm t$, and
defined
$\kappa \!=\! \kappa_a \!=\! \kappa_b$,
$K\!=\! K_a \!=\! K_b$,
$\Delta \!=\! \Delta_a \!=\! \Delta_b$, 
$\Gamma \!=\! \Gamma_a \!=\! \Gamma_b$, $g \!=\! g_a \!=\! g_b$,
and $\Omega \!=\! \Omega_a \!=\! \Omega_b$.
In Fig.~\ref{fig:figure_pra_2} we show the functions
$|\alpha(t)|^2$, $|\beta(t)|^2$, $|\gamma(t)|^2$, 
and $|\delta(t)|^2$, i.e.
the occupation probabilities of the state $|a\rangle$,
$|b\rangle$, $|c\rangle$, and $|d\rangle$, respectively, 
for the case of equal parameters for the
two subsystems $A$ and $B$, with 
$g/K = 5$, $\kappa/K = 0.9$, $\Delta/K = 0.1$, and $\Gamma/K= 0.2$.
For these functions
the phase factor $e^{i \phi}$ does 
not play any role.

\section{Entanglement evolution}
\label{section3}

In the system under study,
the two atoms constitute a pair of qubits.
An appropriate measure of the entanglement
for a two qubits system, often considered in the context of
quantum information theory, is the concurrence~\cite{Wootters:2245}.
Given the density matrix $\rho$ for such a system,
the concurrence is defined as
\begin{equation}
C(\rho) = \max \left\{ 0, \sqrt{\lambda_1} -
\sqrt{\lambda_2}- \sqrt{\lambda_3}- \sqrt{\lambda_4} 
\right\} \, ,
\label{eq:conc}
\end{equation}
where $\lambda_1 \geq  \lambda_2 \geq \lambda_3
\geq \lambda_4$ are the eigenvalues of the matrix
$\tilde \rho = \rho (\sigma_y \otimes \sigma_y)
\rho^{*} (\sigma_y \otimes \sigma_y)$.
Here $\sigma_y$ 
is the Pauli
spin matrix and complex conjugation is denoted by an 
asterisk.
The concurrence varies in the range $[0,1]$,
where the values $0$ and $1$ represent separable states
and maximally entangled states, respectively.

\begin{figure}
\includegraphics[width=7.5cm]{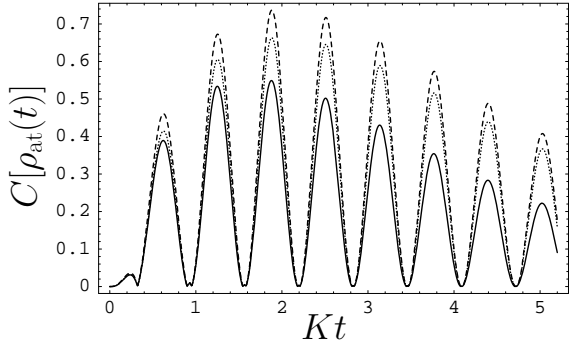}
\caption{The concurrence $C[\rho_{\rm at}(t)]$ between
the two atoms is shown for 
the case of equal parameters for the
two subsystems $A$ and $B$, where $g/K = 5$ and 
$\Delta/K = 0.1$. Moreover, $\kappa/K = 0.9$ and 
$\Gamma/K = 0.2$ (solid line), $\kappa/K = 0.9$
and $\Gamma/K = 0$ (dotted line), $\kappa/K = 1$
and $\Gamma/K = 0$ (dashed line).}
\label{fig:figure_pra_3}
\end{figure} 
To derive an expression for the  
concurrence between the two atoms,
let us consider
the density operator that describes the system.
It is obtained from
the density operator $\hat \rho (t)$, Eq.~(\ref{eq:rho_t_1}),
by tracing over the intracavity field states for 
the two subsystems, $\hat \rho_{\rm at}(t) \!=\! {\rm Tr}_{\rm cav} 
\left[ \hat \rho (t) \right]$,
and is given by
\begin{eqnarray}
\hat \rho_{\rm at}(t) &=&  
|\alpha(t)|^2 |1,0 \rangle \langle 1,0| +
|\gamma(t)|^2 |0,1 \rangle \langle 0,1| \nonumber \\
&+& \alpha(t)\gamma^{*}(t)|1,0 \rangle \langle 0,1|
+\alpha^{*}(t)\gamma(t)|0,1 \rangle \langle 1,0|
\nonumber \\
&+& \left\{ 1 - \left[|\alpha(t)|^2 + |\gamma(t)|^2   
\right]  \right\} |0,0 \rangle \langle 0,0| \, .
\label{eq:rho_atoms}
\end{eqnarray}
Considering the $4\times4$ density matrix 
$\rho_{\rm at}(t)$,
it is easy to show that the concurrence $C[\rho_{\rm at}(t)]$ is, using 
Eq.~(\ref{eq:conc}), given by
\begin{equation}
C[\rho_{\rm at}(t)] = 2 \left| \alpha(t) \right| \left| 
\gamma(t)  \right| \, .
\label{eq:conc_atoms}
\end{equation}
To analyze the time dependence of the concurrence between the two
atoms,
let us consider the case of equal parameters for the
two subsystems $A$ and $B$. Inserting
the analytical solutions 
(\ref{eq:diff_eq_sol_general_a_b}) and (\ref{eq:diff_eq_sol_equal_values})
into Eq.~(\ref{eq:conc_atoms}), we show
in Fig.~\ref{fig:figure_pra_3}
the function $C[\rho_{\rm at}(t)]$
for the parameters $g/K = 5$, 
$\Delta/K = 0.1$,
and for different values of $\kappa/K$ and $\Gamma/K$.
Since the concurrence contains only
absolute values, the phase 
factor $e^{i \phi}$ does not play any role here.

From Fig.~\ref{fig:figure_pra_3} one can clearly see that the
initially disentangled atoms become
entangled. In particular,
a maximum value for $C[\rho_{\rm at}(t)]$ is
found for $\bar t \simeq 1.88/K$, where,
for the shown cases,
$C[\rho_{\rm at}(\bar t)] \simeq 0.74,~0.66$, and $0.55$.
>From the entanglement between the two atoms, as  
shown in Fig.~\ref{fig:figure_pra_3}, it is clearly seen
that it sensitively depends on unwanted losses.
Already the effects due to the absorption or scattering
by the cavity mirrors alone are not negligible.
Moreover, the inclusion of the
spontaneous emission further reduces the degree of the 
entanglement between the two atoms. 
For example, the relative variation of
the concurrence is approximately $10\%$ between
the ideal case $\kappa/K=1$ and $\Gamma/K=0$
(no absorption or scattering and no spontaneous emission) and the case $\kappa/K=0.9$ and $\Gamma/K=0$,
considering the peak at $\bar t$. The concurrence
decreases even further if spontaneous emissions are
included. In this case for $\kappa/K=0.9$ and $\Gamma/K=0.2$ 
the peak of the concurrence at $\bar t$ is reduced
by approximately $25\%$ compared with the ideal lossless case.
These values clearly tell us that the effects of unwanted losses
cannot be neglected in general, when considering the entanglement
in realistic quantum systems.
Of course, for $K t \gg 1$, the two atoms become 
again disentangled due to the emission, sooner or later,
of a photon into one of the five decay channels.
The release of a photon into the environment destroys 
any entanglement, projecting the two-atom subsystem 
onto the separable state $|0,0\rangle$.

\begin{figure}
\includegraphics[width=7.5cm]{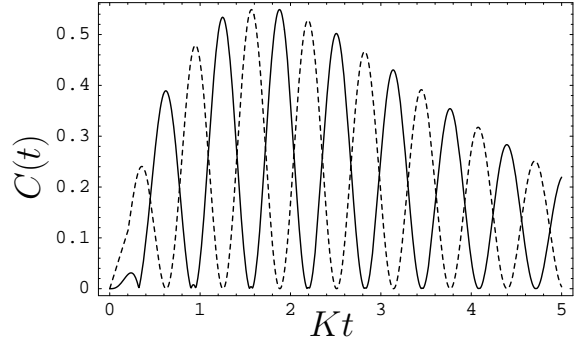}
\caption{Comparison between the concurrences for the two atoms (solid line) and for the two intracavity fields (dashed line).
The parameters for the
two subsystems are equal, with 
$g/K = 5$, $\Delta/K = 0.1$, $\kappa/K = 0.9$ and 
$\Gamma/K = 0.2$.}
\label{fig:figure_pra_4}
\end{figure} 
Finally, we note that the concurrence between
the two intracavity fields can be obtained as well.
Let us consider 
the density operator that describes the system
of the two intracavity fields $A$ and $B$,
obtained from the density operator $\hat \rho (t)$, cf. Eq.~(\ref{eq:rho_t_1}), by tracing over the atomic states
of the two subsystems,
$\hat \rho_{\rm cav}(t) \!=\! {\rm Tr}_{\rm at} 
\left[ \hat \rho (t) \right]$.
Considering now the $4\times4$ density matrix 
$\rho_{\rm cav}(t)$ 
in the two intracavity-field Fock basis, the concurrence 
$C[\rho_{\rm cav}(t)]$ is given by
\begin{equation}
C[\rho_{\rm cav}(t)] = 2 \left| \beta(t) \right| \left| 
\delta(t)  \right| \, .
\label{eq:conc_cavities}
\end{equation}
A comparison between the concurrence for the two atoms and the
concurrence for the two intracavity fields is shown in Fig.~\ref{fig:figure_pra_4}. 
When the concurrence
between the two atoms reaches a maximum, the concurrence
between the two intracavity fields is approximately zero,
and vice  versa.
This is related to the fact that the 
excitation energy is transferred between the atoms and the 
intracavity fields.
When the atoms are unexcited, their state is separable and the intracavity
fields are entangled. As a function of time, the entanglement is thus exchanged between the two atoms and the two intracavity fields.

\section{Summary}
\label{conclusions}

The dynamics of a cascaded system that
consists of two atom-cavity subsystems has
been analyzed. 
Unwanted losses have been included, such as photon absorption
and scattering by the cavity mirrors and spontaneous emission
of the atoms.
The evolution of
the open quantum system under study has been described 
by means of a master equation. 
Considering an initially excited two-level atom in 
the source subsystem, analytical solutions for the dynamics of the system
have been obtained.
The entanglement evolution between the two atoms,
constituting a two-qubit system,
has been studied by using the concurrence.
A similar analysis has been performed for the 
two intracavity fields.

The dynamical evolution of the system shows that 
the two initially disentangled qubits reach states
of significant entanglement.
It has been shown that the
entanglement generated between the two atoms 
sensitively diminishes due to the presence of unwanted losses,
which cannot be neglected in realistic quantum systems.
We have also shown that the entanglement is periodically exchanged
between the two atoms and the two intracavity fields.

\section*{Acknowledgments}
This work was supported by the Deutsche Forschungsgemeinschaft.

\end{document}